\documentclass[%
 reprint,
superscriptaddress,
showkeys,
 amsmath,amssymb,
 aps,
prx, 
]{revtex4-2}

\usepackage{graphicx}
\usepackage[caption=false]{subfig} 

\usepackage{dcolumn}
\usepackage{bm}

\newcommand{\ee}{\end{equation}} 
\newcommand{\be}{\begin{equation}}

\usepackage[mathscr]{euscript}  
\usepackage{xcolor}  

\usepackage{tcolorbox}
\usepackage{comment}
\usepackage{float}

\makeatletter
\newsavebox{\@brx}
\newcommand{\llangle}[1][]{\savebox{\@brx}{\(\m@th{#1\langle}\)}%
  \mathopen{\copy\@brx\kern-0.5\wd\@brx\usebox{\@brx}}}
\newcommand{\rrangle}[1][]{\savebox{\@brx}{\(\m@th{#1\rangle}\)}%
  \mathclose{\copy\@brx\kern-0.5\wd\@brx\usebox{\@brx}}}
\makeatother

\begin{document}

\preprint{APS/123-QED}

\title{Trapping of active Brownian and run-and-tumble particles:\\ a first-passage  time approach}
 
\author{Emily Qing Zang Moen}
\affiliation{The Njord Center, Department of Physics, University of Oslo, Blindern, 0316 Oslo, Norway\\}

\author{Kristian St\o{}levik Olsen}
\affiliation{Nordita, Royal Institute of Technology and Stockholm University,
Hannes Alfvéns väg 12, 23, SE-106 91 Stockholm, Sweden\\}

\author{Jonas R\o{}nning}
\affiliation{The Njord Center, Department of Physics, University of Oslo, Blindern, 0316 Oslo, Norway\\}

\author{Luiza Angheluta}
\affiliation{The Njord Center, Department of Physics, University of Oslo, Blindern, 0316 Oslo, Norway\\}

\date{\today}

\begin{abstract}
We use a first-passage time approach to study the statistics of the trapping times induced by persistent motion of active particles colliding with flat boundaries. The angular first-passage time distribution and mean first-passage time is calculated exactly for active Brownian and run-and-tumble particles and the results are compared. Theoretical predictions are in excellent agreement with Langevin simulations. Our results shed further light onto how active particles with different dynamics may be equivalent in the bulk, yet behave differently near boundaries or obstacles. 
\end{abstract}

\pacs{Valid PACS appear here} 
\maketitle

\section{Introduction}

Active matter is a class of non-equilibrium systems where the constituents are self-propelled by consuming energy from their environment. Much of the theoretical interest in the field comes from the diverse emergent phenomena with no counterpart in analogous equilibrium or passive systems. The role of disorder and confinement has gained particular attention since these are present in most realistic active matter systems, like those found in biological matter, but also due to potential microfluidic applications. A myriad of studies have revealed the importance of boundaries in shaping the macroscopic behavior of active systems \cite{dor2021far}, including novel collective states in solid confinement \cite{jiang2017emergence,olsen2020escape}, and surface accumulation and currents \cite{caprini2019active,lee2013active,elgeti2013wall}.

\begin{figure}
    \centering
    \includegraphics[width = 8.5 cm]{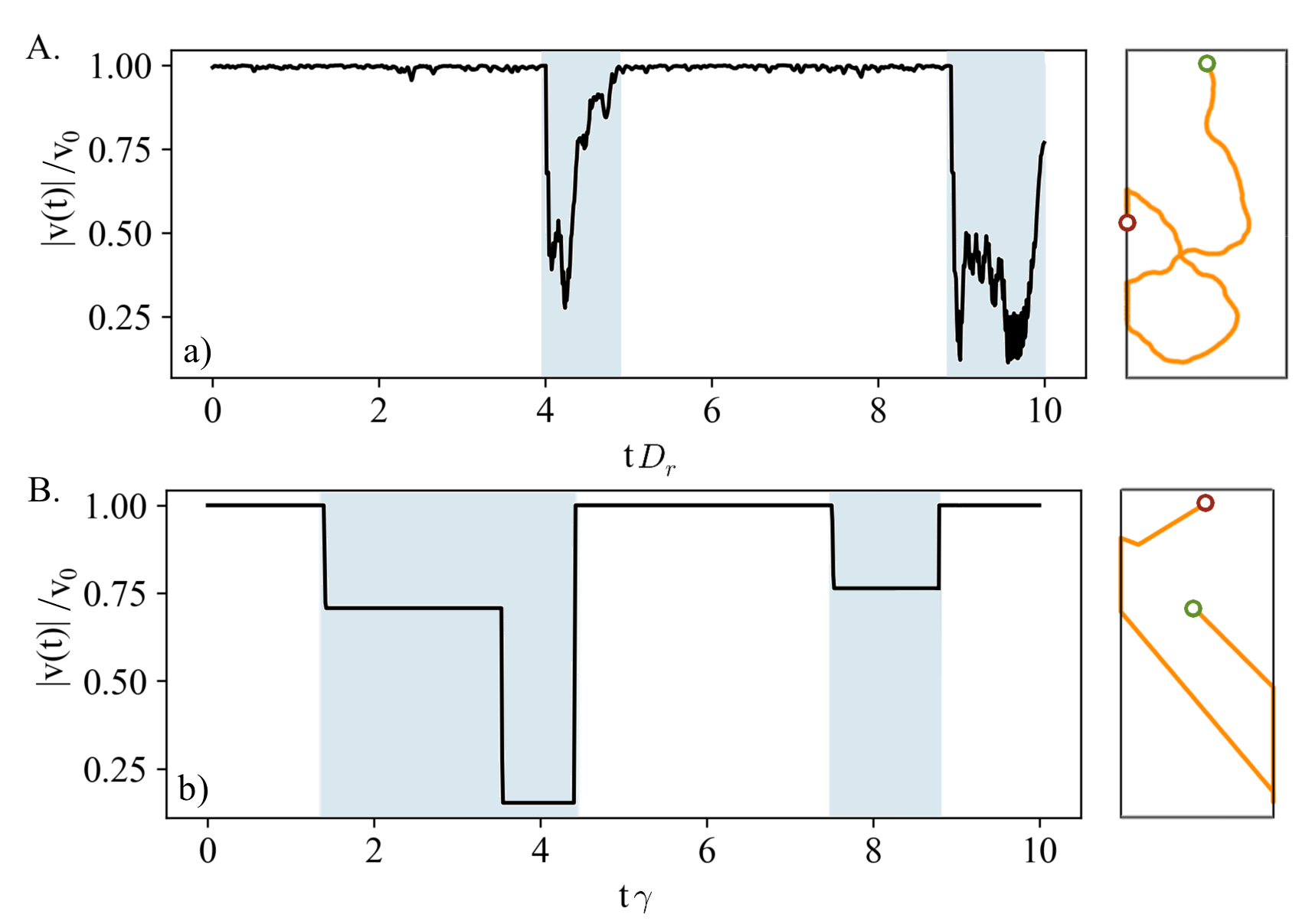}
    \caption{Sketch of the system under consideration, where active Brownian particles (A) and run-and-tumble particles (B) collide and get trapped at solid flat boundaries (rightmost figures). Speed time-series shows trapping times as periods of reduced self-propulsion compared to bulk motion. Trapping durations are indicated with blue shaded regions. }
    \label{fig:fig1}
\end{figure}

In this paper, we study the statistical features of trapping induced by persistent motion of active particles near a flat, solid boundary. As limit cases, we
compare the statistics of trapping times corresponding to active Brownian particles (ABPs) and run-and-tumble particles (RTPs). While these two minimal models are well-known to give rise to the same statistical properties in open spaces \cite{cates2013active}, we show through analytical calculations and numerical simulations that the two  differ significantly under confinement where the microscopic details of persistent motion become important for the  wall-trapping statistics. This is of relevance to phenomena like surface accumulation, but also for example in studied of pressure  in active systems which will depend on particle-boundary contact times. Previous works have also reported different behavior of ABPs and RTPs in confinement. For example, in a harmonic confinement, the ABPs tend to drift more around the high-potential regions of the potential than RTPs \cite{solon2015active}. It has also been shown that the two particle types separate when released inside a maze-like geometry \cite{khatami2016active}. This separation was also attributed to difference in surface interactions. The surface residence time was studied for a modified run-and-tumble model and compared with E-coli data in Ref.~\cite{junot2021run}, and the effect of hydrodynamic interactions on the residence times of ABPs were studied in Ref.~\cite{schaar2015detention}.
However, there is a subtlety in how the residence time is defined and how it impacts the residence time statistics. In previous works, the residence time was typically calculated by introducing a boundary layer of an adhoc width and measuring the time spend by the active particle within that layer. We here propose an alternative way of achieving the trapping time, suitable for the case of simple steric body forces between particle and boundary. Namely, we define the trapping as the time that the particle \emph{effectively} exchanges momentum with the wall. In other words, here we consider the trapping time as the time that the particle resides at the wall with a non-zero velocity normal to the wall. Fig. (\ref{fig:fig1}) illustrates trapping events of both ABPs and RTPs in a box.

In a minimal stochastic model, the 2D dynamics of an active particle is typically described by following stochastic equations \cite{fodor2018statistical}
\begin{align}\label{eq:dyneqs}
    \dot {\mathbf{x}} &= v_0 \hat{\mathbf{e}}(t) + \mathbf{F}_c\\
    \dot \theta &= \sqrt{2 D_r} \xi(t) +  \sum_{\alpha}\Delta \theta_\alpha \delta(t-t_\alpha).
\end{align}
Here $v_0$ is the constant, self-propulsion speed, $D_r$ is the rotational diffusivity, and $t_\alpha$ are random tumbling times generated by a Poisson process with rate $\gamma$. Here $\Delta \theta_\alpha$ are uniform random angles describing the tumbling angle.  The case $D_r = 0$ yields pure run-and-tumble particles, while $\gamma = 0$ results in active Brownian motion. These are the two cases that we consider separately. The confinement force $\mathbf{F}_c$ is taken to be a purely steric body force modelled as
\begin{equation}
\mathbf{F}_c = 
\left\{
	\begin{array}{ll}
		 - v_0 (\hat{\mathbf{e}}\cdot \hat{\mathbf{n}}) \hat{\mathbf{n}}  & \mbox{at boundary}  \\
		0 & \mbox{in bulk.}
	\end{array}
\right.
\end{equation}
where $\hat{\mathbf{n}}$ is the outward unit vector of the wall boundary. Such steric boundary interactions do not induce any torques on the particle's direction of motion, so that the angular dynamics remains the same both at the boundary and in the bulk. Additional boundary effects on the angular dynamics may be realistic if hydrodynamic interactions are taken into account, or in the particles in questions have a non-spherical elongated shape \cite{elgeti2016microswimmers,elgeti2009self}. Different kinds of migrating cells or swimming bacteria display various scattering behavior from surfaces\cite{kantsler2013ciliary,lushi2017scattering}. We focus primarily on the effect of persistent motion into flat walls of pointwise active particles and aim to understand the effect of the two different types of angular dynamics on wall-trapping statistics. 

Our approach is to map the problem of finding trapping duration into a first-passage problem for the marginalized angular dynamics for active Brownians ($\gamma=0$) and run-and-tumble particles ($D_r=0$), respectively. This angular first-passage time (FPT) distribution is calculated straightforwardly from the relation to the survival probability $S (t,\theta_a | \theta_0)$. This is the probability that the angular stochastic process remains inside an angular domain $(-\theta_a,\theta_a)$ up to time $t$ given an initial angle $\theta_0$.  By solving the corresponding Fokker-Planck or master equation for the angular distributions $\rho_a(\theta,t | \theta_0)$ conditioned on some incoming angle $\theta_0$ and with absorbing boundary conditions at the two target angles, we can then compute the survival probability by 
\begin{equation}\label{eq:s1}
S (t,\theta_a | \theta_0)  = \int_{-\theta_a}^{\theta_a} d\theta \rho_a(\theta,t | \theta_0),    
\end{equation}
which is also conditioned on the incoming angle. 
The FPT distribution follows then as  \cite{redner2001guide}
\begin{equation}\label{eq:fptd1}
 F(t, \theta_a |  \theta_0)  = - \frac{d}{dt} S (t,\theta_a | \theta_0). 
\end{equation}
In more realistic situations where a particle encounters a boundary, the collision angle or incoming angle $\theta_0$ will be random, meaning we should interpret $S (t,\theta_a | \theta_0) $ as a conditional probability on the random incoming angle. For a given the distribution $\mathcal{P}(\theta_0)$ of incoming angles, we may integrate out this random variable, resulting in the averaged FPT distribution as 
\begin{equation}\label{eq:fptd2}
\overline{F}(t, \theta_a ) = \int_{- \theta_a}^{ \theta_a} d\theta_0  F(t, \theta_a |  \theta_0)  \mathcal{P}(\theta_0)
\end{equation}
Similarly, we may calculate the conditional mean FPT $T_1(\theta_a|\theta_0) $ and then integrate over the incoming angles to obtain 
\begin{align}\label{eq:mfpt}
\overline{T}_1(\theta_a) &= \int_{- \theta_a}^{ \theta_a} d\theta_0  T_1(\theta_a|\theta_0)   \mathcal{P}(\theta_0) \\
&= \int_{0}^\infty d t \overline{F}(t,\theta_a ) t
\end{align}
From the perspective of the first-passage problem, the incoming angle of a particle when colliding with the boundary is in essence an initial condition, and we may refer to it as such in the following. We however note that it is the same as the initial conditions of the particle itself, which we always initialize far from the boundary.

The types of averages discussed above may dramatically alter the analytical form of the FPT distribution, as observed recently in the first-passage problem of a Brownian process with a single target and normally distributed initial conditions \cite{PhysRevE.104.L012102}. Here it was observed that as the width of the distribution of initial conditions was varied the FPT distribution underwent a transition from monotonic to non-monotonic at a certain critical width. In the case of ABPs, where the angular dynamics is Brownian, a distribution of incoming angles can significantly change the FPT distribution. As we will discuss below, however, this is not the case for RTPs.

The remainder of this article is structured as follows. In section \ref{sec:rtp}, we consider the relevant class of first-passage problems for the rotational dynamics of a run-and-tumble particle, namely the statistics of the first passage out of the angular domain $(-\theta_a,\theta_a)$. The case of trapping at a flat wall is obtained by setting $\theta_a = \pi/2$.  Simple arguments based only on Poissonian nature of tumbles leads to a simple analytical expression for the trapping time distribution in both two and three dimensions. Furthermore, the first-passage time distribution and the corresponding mean have no dependence on the incoming angle of the particles, and are hence insensitive to the distribution of incoming angles.  Section  \ref{sec:abp} considers the active Brownian case in two dimensions. As the marginalized angular dynamics is Brownian, an exact analytical solution is readily obtainable. Upon averaging over incoming angles, various distribution shapes can be found. Section \ref{sec:chan} discusses ABPs and RTPs confined to a 2D infinite channel. Comparisons of the analytical results and numerical simulations are made, showing an excellent agreement. Section \ref{sec:outlook}  provides concluding discussions and potential outlooks.

\section{The first-passage problem}
\begin{figure}
    \centering
    \includegraphics[width = 8.5 cm]{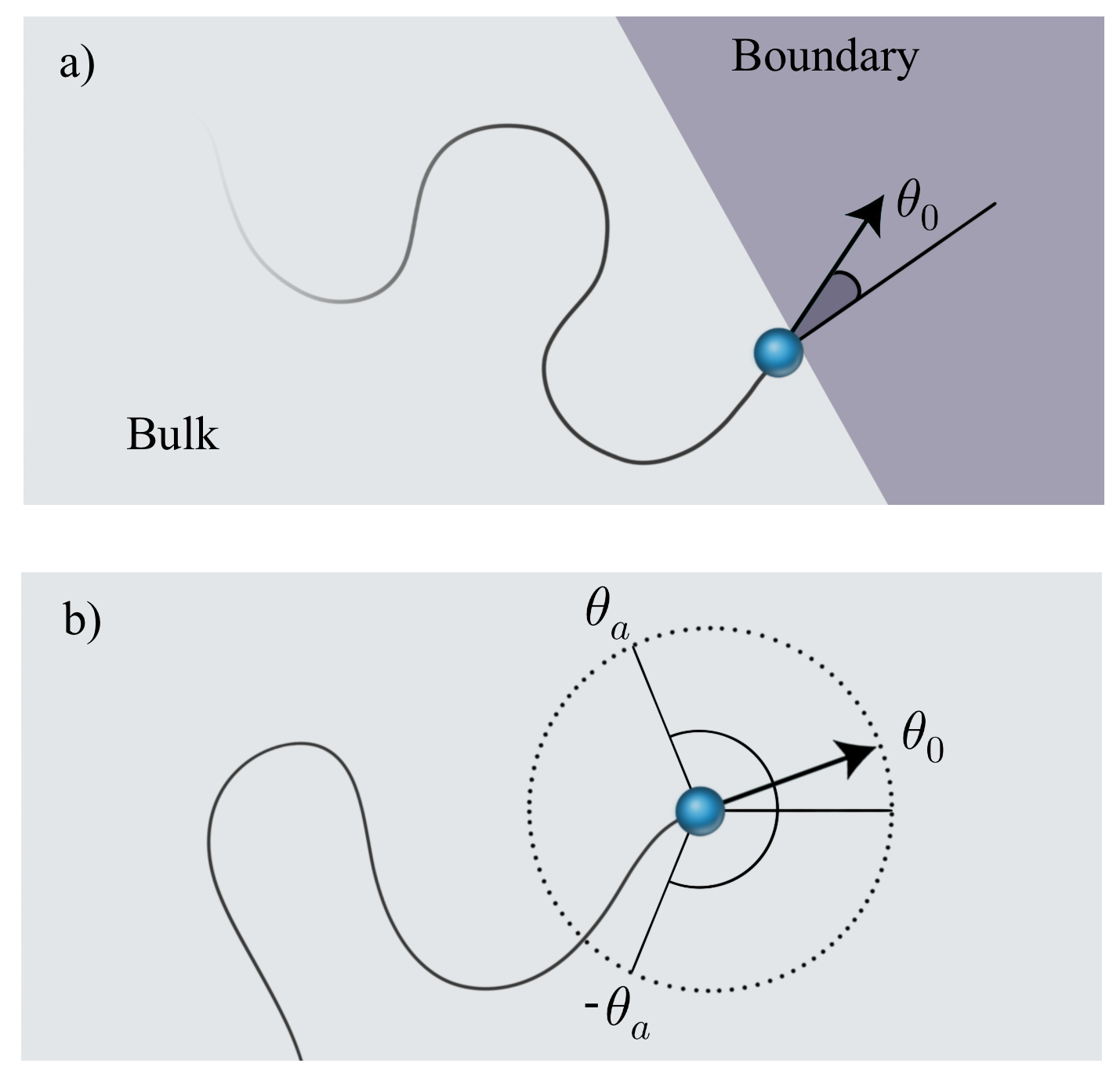}
    \caption{a) Active particles colliding with a flat boundary with an incoming angle $\theta_0$. b) General angular first-passage problem with arbitrary target angle $\theta_a$. }
    \label{fig:2}
\end{figure}

The class of first-passage problems we are considering is based on the survival of a stochastic angular variable in the angular domain $(-\theta_a,\theta_a)$, with absorbing boundary point $\theta_a \leq \pi$. As discussed above, the case $\theta_a = \pi/2$ corresponds to the escape from a flat wall. In this setup, we use a convention so that a particle that collides head-on with a boundary has an incoming angle $\theta_0 = 0$. See Fig. (\ref{fig:2}) for illustration showing the conventions. We first solve the problem for run-and-tumble particles, before proceeding to the active Brownian case.

\subsection{Run-and-tumble case}\label{sec:rtp}
For run-and-tumble particles, the angular variable performs discontinuous jumps on the circle at random times that obey Poisson statistics. This is therefore equivalent to a discrete random walk on the circle with arbitrarily large (but periodic) step sizes. Moreover, the absorbing targets are not point sinks but rather extended regions, similar to  the recent study in Ref.~\cite{pozzoli2021survival}.

\begin{figure}[t ]
    \centering
    \includegraphics[width = 8.5cm]{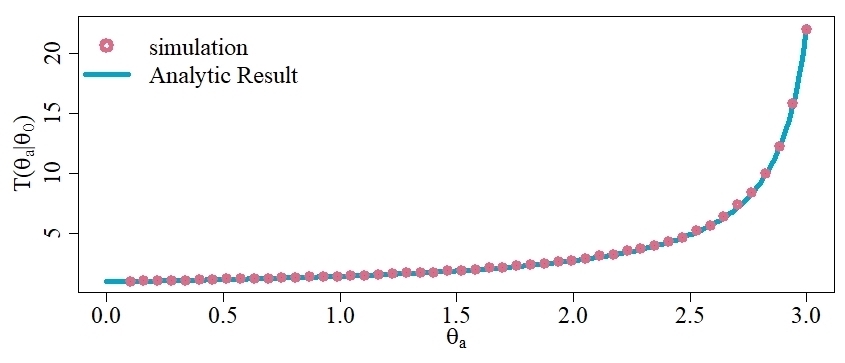}
    \caption{Plot showing the mean FPT  in units of the persistence time as a function of target angle $\theta_a$ for the RTP case, showing a clear agreement between theory and simulations.} \label{fig:3}
\end{figure}

To calculate the survival probability, we assume that in the interval $(0,t)$ there has been exactly $N_{tum}(t) = n$ tumbles
\begin{equation}
   \theta_0 \to  \theta_1\to\theta_2\to\theta_3\to ... \to \theta_n.
\end{equation}
The probability that this has happened is simply given by the Poisson probability $ P[N_{tum}(t) = n]  = \exp(-\gamma t) (\gamma t)^n /n!$. To find the survival probability that the angle remains inside a domain $(-\theta_a,\theta_a)$ during these tumbles, we note that the tumbles are independent and equivalent to uniformly picking a random point on the circle. Hence the survival probability conditioned on a fixed number of tumbles is simply
\begin{equation}
    S_n(t,\theta_a|\theta_0) =\Theta(\theta_a - |\theta_0|) P[N_{tum}(t) = n]  \left(\frac{\theta_a}{\pi}\right)^n
\end{equation}
Here $\theta_a/\pi$ is the probability that a single tumble remains in $(-\theta_a,\theta_a)$, and the Heaviside theta function ensures that the survival probability is zero if the initial angle is outside the domain of interest. The survival probability in time $t$ is then obtained by marginalizing over the number of tumbles
\begin{align}
    S(t,\theta_a|\theta_0) &= \sum_{n=0}^\infty S_n(t,\theta_a|\theta_0) \\
    &= \Theta(\theta_a - |\theta_0|) e^{-\gamma\left(1 - \frac{\theta_a}{\pi}\right)t}
\end{align}
From Eq. (\ref{eq:fptd1}) we can easily calculate the FPT distribution
\begin{equation}
    F(t,\theta_a|\theta_0) =  \Theta(\theta_a - |\theta_0|)\gamma\left(1 - \frac{\theta_a}{\pi}\right) e^{-\gamma\left(1 - \frac{\theta_a}{\pi}\right)t}
\end{equation}
 and its mean as
\begin{equation}
    T_1(\theta_a|\theta_0) =  \frac{\Theta(\theta_a - |\theta_0|)}{\gamma\left(1 - \frac{\theta_a}{\pi}\right)}
\end{equation}
Interestingly, the mean has a divergence as $\theta_a \to \pi$. This makes sense, as the probability of tumbling out of the domain $(-\theta_a,\theta_a)$ vanishes in this limit, and hence we should expect an infinite first-passage time. To verify this result, we perform stochastic simulations of the tumbling dynamics and show in Fig. (\ref{fig:3}) that the dependence on $\theta_a$ of the mean first-passage time is in excellent agreement with the theoretical prediction.

The case corresponding to trapping at a flat boundary $\theta_a = \pi/2$ results in 
\begin{equation}
    F(t,\pi/2|\theta_0)  =  \Theta\left(\frac{\pi}{2} - |\theta_0|\right) \frac{\gamma}{2} e^{-\gamma t/ 2}
\end{equation}
Assuming that $|\theta_0| < \theta_a$, we can ignore the Heaviside step function in the above expressions. In this case there is no remaining dependence on the collision angle $\theta_0$, and hence we have trivially that
\begin{equation}
    \overline{F}(t,\theta_a) = F(t,\theta_a| \theta_0)
\end{equation}
and similarly for the mean FPT. This behavior originates in the fact that run-and-tumble dynamics is described by uniform transition probabilities. In the more general case, where tumbling angles follow a distribution, we expect that the collision angle plays a more prominent role. 

The above arguments generalize trivially to any stochastic search process where space is sampled randomly and uniformly at times following a Poisson process. In particular this means that in the case of three dimensions, where the direction of motion is given by a point on a unit sphere, the FPT distribution is similarly given as 
\begin{equation}
    F_{3d}(t,\mathcal{D}|\theta_0) =  \gamma\left(1 - \frac{|\mathcal{D}|}{4 \pi}\right) e^{-\gamma\left(1 - \frac{|\mathcal{D}|}{4 \pi}\right)t}
\end{equation}
where $\mathcal{D}\subset \mathbf{S}^2$ is the domain on the unit sphere out of which we are interested in tumbling, and $4\pi$ is the spheres total area. When an active particle is trapped at a 2-dimensional boundary, the escape takes place when the director leaves the upper half-sphere. In this case $|\mathcal{D}| = 2\pi$, and the FPT distribution reads as
\begin{equation}
    \overline{F}_{3d}(t) =  \frac{\gamma}{2} e^{-\gamma t/ 2}
\end{equation}
The same argument extends to a $d$-dimensional particle trapped at a $(d-1)$-dimensional hypersurface, and in this sense the distribution is universal in $d\geq 2$. Alternatively, one can obtain the new direction of motion after a tumble by picking a uniformly random point on a $d$-dimensional sphere. This is typically done by generating a length $d$ vector of standard normal variables. Upon normalization this vector points to a uniformly random point on the sphere. A RTP will be able to escape from the wall when \emph{a single} one of these $d$ normal variables changes time, which happens with rate $\gamma/2$ when tumbles take place at rate $\gamma$. Hence the above FPT distribution is universal for all dimensions, and we will refer to it simply as $\overline{F}(t)$.


\subsection{Active Brownian case}\label{sec:abp}
In contrast to the run-and-tumble case, the angular dynamics of an active Brownian particle is continuous and Brownian, and can be solved using standard Fokker-Planck methods. The full dynamics of the distribution $P (\vec{x},\theta,t)$ is in general described by a Fokker-Planck equation of the form 
\begin{equation}
    \partial_t P  + v_0 [\hat e (\theta) \cdot \nabla_x] P = D_r\partial_\theta^2 P 
\end{equation}
where $\hat e (\theta) = [\cos\theta,\sin\theta]$ is the unit vector in the direction of motion of motion of the particles in which they move with constant speed $v_0$. Marginalizing over the spatial variable $\rho = \int d \vec{x} P$ results simply in the diffusion equation $\partial_t \rho = D_r \partial_\theta^2 \rho$. Since the process is terminated before the angles have explored the full circle, no effects of periodicity are present in this calculation and the problem is equivalent to that of normal Brownian motion on an interval.  This classical problem was first studied in the 50s by Darling and Siegert and has since then appeared in multiple references with various generalizations \cite{darling1953first,redner2001guide,bray2013persistence, pal2019first,pal2019motion,redner2022look}. A standard approach to obtaining first-passage times for such processes is to find the Laplace transform of the survival probability by means of a backward Fokker-Planck equation. From this the mean FPT is easily obtained. Another approach uses the well-known fact that for Fokker-Planck equations of the type $\partial_t P = \mathcal{L}_\text{FP} P$ the mean FPT satisfies $\mathcal{L}_\text{FP} ^\dagger  T_1 = -1$, evaluated at the initial condition \cite{zwanzig2001nonequilibrium}. In the diffusive case this is simply a boundary value problem for the Poisson equation. In either case, the mean FPT is known to take the form \cite{redner2001guide, redner2022look}
\begin{equation}\label{eq:classic}
    T_1(\theta_a|\theta_0) = \frac{\theta_a^2 - \theta_0^2}{2 D_r}.
\end{equation}
The aforementioned approaches, either by means of Laplace transforms of the Poisson equation, allows one to easily obtain the mean FPT directly from the Fokker-Planck equation. However, here we are also interested in the analytical form of the first-passage time distribution. We therefore directly solve the Fokker-Planck equation by means of an eigenfunction expansion, and derive an expression for the mean FPT conditioned on an initial angle. This expression may then be used when we average over the distribution of initial angles. 

We solve the rotational diffusion problem 
\begin{equation}
    \partial_{t} \rho_a(\theta,t) = D_r \partial_{ \theta}^2\rho_a(\theta,t), \rho_a(\pm \theta_a,t) = 0
\end{equation}
with initial condition a Dirac delta function $\rho_a(\theta,0) = \delta(\theta-\theta_0)$ located at an initial angle $\theta_0$ that, like in the run-and-tumble case, is assumed to lie somewhere between the two absorbing angles. The solution to the above boundary value problem can be expressed as a series
\begin{align}
    \rho_a(\theta,t) &= \sum_{n=1}^\infty a_n(t) \psi_n(\theta)\\
    &= \sum_{n=1}^\infty a_n(t) \sin\left(\frac{n\pi(\theta-\theta_a)}{2 \theta_a}\right)
\end{align}
where $\psi_n(\theta)$ are eigenfunctions of the Dirichlet Laplacian forming a complete basis for function on the domain $(-\theta_a,\theta_a)$. Insertion into the diffusion equation gives the time evolution of the time-dependent coefficients as $a_n(t) = a_n(0) \exp(- D_r (n\pi/2\theta_a)^2 t)$. Through the orthogonality of the eigenfunctions we can find the remaining coefficients $a_n(0)$ from the initial condition as 
\begin{equation}
    a_n(0) = \frac{1}{\theta_a} \int_{-\theta_a}^{\theta_a} d\theta \rho_a(\theta,0)\psi_n(\theta)
\end{equation}
For a Dirac delta initial condition located at $\delta(\theta-\theta_0)$, the initial coefficients are simply $a_n(0) = \psi_n(\theta_0)/\theta_a$, giving a solution
\begin{equation}\label{eq:FPsolution}
    \rho_a(\theta,t) = \frac{1}{\theta_a}\sum_{n=1}^\infty e^{-D_r\left(\frac{n\pi}{2\theta_a}\right)^2 t} \psi_n(\theta_0) \psi_n(\theta) 
\end{equation}
In the case of a Brownian process on the real line with a single absorbing target, the image method is a popular method used to satisfy the boundary conditions. The present case can also be solved through this method, by appropriately placing image densities on either sides of the absorbing angles $\pm\theta_a$. For each of these image densities to have the correct behavior, they will again need image densities of their own. This leads to an infinite sum over densities and their images, which is equivalent to the above solution.

To obtain the survival probability, we use Eq. (\ref{eq:s1}) together with the solution in Eq. (\ref{eq:FPsolution}). When integrating over the domain $(-\theta_a,\theta_a)$ we first note that the even terms in Eq. (\ref{eq:FPsolution}) will vanish, since $\psi_{2n}(\theta)$ are functions of odd parity. Integrating the remaining odd terms in Eq. (\ref{eq:FPsolution}) results in the survival probability
\begin{equation}
	S(t, \theta_a | \theta_0)  = \frac{2}{ \theta_a }\sum_{n = 0}^{\infty}(-1)^n \frac{1}{\mu_n}\cos(\mu_n \theta_0)  e^{-D_r \mu_n^2 t}
\end{equation}
where we have introduced $\mu_n = (2n+1)\pi /(2\theta_a)$ for notational simplicity. The corresponding first-passage time distribution is
\begin{align}\label{eq:Hcondinitial}
	F(t, \theta_a |  \theta_0)  & 	=  \frac{2D_r}{\theta_a}\sum_{n = 0}^{\infty}(-1)^n   \mu_n \cos(\mu_n \theta_0)  e^{-D_r \mu_n^2 t} \\
& 	= \frac{D_r}{\theta_a} \frac{d}{d\theta_0} \vartheta_1\left(e^{-D_r \pi^2 t /\theta_a^2}, \frac{\pi \theta_0}{2\theta_a}\right) 
\end{align}
where we used the elliptic theta $\vartheta_1(q,z) = 2 \sum_{n=0}^\infty (-1)^n q^{(n+1/2)^2} \sin\left[ (2n+1) z\right]$, see eg. Ref.~\cite{whittaker2020course}. Since $\lim_{t\to \infty}S(t,\theta_a|\theta_0) = 0$, we also have that 
\begin{align}\label{eq:norm}
    \int_0^\infty dt F(t, \theta_a |  \theta_0) &= S(0,\theta_a|\theta_0) \\
    &= \frac{2}{\theta_a}\sum_{n = 0}^{\infty}(-1)^n \frac{\cos(\mu_n \theta_0)  }{\mu_n}
\end{align}
which takes a form similar to the series expansion of the arctangent. Expressing the cosine as the sum of complex exponentials and using the identity $\tan^{-1}(e^{ix})+ \tan^{-1}(e^{-ix}) = \pi/2$ for $|x|<\pi/2$ one can verify that the sum evaluates to unity for all $\theta_0$.

Now, we want to check that our FPT distribution has a mean with the same quadratic dependence on $\theta_a$ as in Eq. (\ref{eq:classic}). The mean FPT as function of $\theta_a$ and conditioned on the incoming angle $\theta_0$ can be calculated directly as the first moment of the FPT distribution, namely 
\begin{align}\label{eq:mfpt1}
   T_1(\theta_a|\theta_0)  &=  \int_0^\infty dt t F(t, \theta_a |  \theta_0)  \\
   &= \frac{2}{D_r \theta_a}\sum_{n = 0}^{\infty}(-1)^n \frac{\cos(\mu_n \theta_0)  }{\mu_n^3}
\end{align}
which is an analytical function of $\theta_0$, and thus has a well-defined Taylor series expansion given as 
$$T_1(\theta_a|\theta_0)   = \sum_{k=0}^\infty\left \{
    \frac{2}{D_r \theta_a} \sum_{n = 0}^{\infty} (-1)^{n+k} \frac{\mu_n ^{2k-3}}{(2k)!} \right\}\theta_0^{2k}$$
Here it is worth noting that the series representation of the Taylor coefficients, given by the expression in the brackets, are divergent for $k\geq 2$ since the $\mu_n$ grows linearly in $n$. However, regularization schemes like Abel summation can be used to assign to all these higher order coefficients the value zero, leaving only the first two terms non-zero. This results in the well-known expression 
\begin{equation}\label{eq:mfptmain}
	 T_1(\theta_a|\theta_0) = \frac{\theta_a^2 - \theta_0^2}{2 D_r},
\end{equation} 
in agreement with the classical result Eq. (\ref{eq:classic}) as expected.

When averaging over a distribution $\mathcal{P}(\theta_0)$ of initial angles, we note that the FPT distribution takes the general form 
\begin{align}
     & \overline{F}(t, \theta_a )  	=   \nonumber \\ &\frac{2D_r}{\theta_a}\sum_{n = 0}^{\infty}(-1)^n   \mu_n  \left(\int d\theta_0 \mathcal{P}(\theta_0)\cos(\mu_n \theta_0) \right)e^{-D_r \mu_n^2 t}
\end{align}
To proceed, we note that the collision angles $\theta_0$ when an active particle encounters a flat boundary can never be at $\pm \pi/2$, as this corresponds to motion paralell to the boundary. In the context of the present first-passage problem with general target angle $\theta_a$, this corresponds to imposing zeros $\mathcal{P}(\pm \theta_a) =0$. Since the particles are achiral we also expect by symmetry that $\mathcal{P}$ is an even function. Combined, these facts imply that we can write $\mathcal{P}$ as the cosine series
\begin{equation}
    \mathcal{P}(\theta_0) = \sum_n \tilde{ \mathcal{P}}_n \cos(\mu_n \theta_0).
\end{equation}
In this case, we can again use orthogonality to write
\begin{equation}
    \int d\theta_0 \mathcal{P}(\theta_0)\cos(\mu_n \theta_0) = \theta_a \tilde{\mathcal{P}}_n
\end{equation}
which leads to the final expression for the collision-averaged FPDT distribution
\begin{equation}
    \overline{F}(t, \theta_a )  	=  2 D_r\sum_{n = 0}^{\infty}(-1)^n   \mu_n  \tilde{\mathcal{P}}_n e^{-D_r \mu_n^2 t}.
    \label{eq:avgFPTD}
\end{equation}
It is worth noting that averaging over collisions will not affect the tail of the distribution $\overline{F}\sim \exp(- D_r t)$, but rather influences only its behavior at short trapping times. This is sensible also intuitively, as the particles that contribute to the tail of the distribution are those that spend a long time finding any of the absorbing targets, and hence no longer remember their initial collision angle. The short passage times are determined by the collision angle distribution, and by changing the coefficients $\tilde{\mathcal{P}}_n$ one can observe both monotonic and non-monotonic behaviors in $\overline{F}$. Different distributions $\mathcal{P}$ can in principle be constructed by experimentally or numerically considering various setups for releasing the particles near a flat boundary. In addition, we can find the averaged mean FPT as
\begin{equation}
	\overline{T}(\theta_a)  =  \frac{2}{D_r}\sum_{n= 0}^{\infty} 
	  \frac{(-1)^k \tilde{ \mathcal{P}}_n}{\mu_k^3}.
\end{equation}
In the next section we will apply the analytical results obtained in this section to the case of active particles confined to a channel.

\section{ Active particles in a channel}\label{sec:chan}
To verify the above analytical predictions, we consider now numerical simulations of particles confined to a channel, in which case we use $\theta_a = \pi /2$ in the above results. This case of a channel is practical, since the two boundaries in this case prevents the particles from straying too far from the boundaries, which would made accumulating data slow. If the particles are initialized close to one of the two boundaries, memory effects from the initial conditions of the particle may bias the simulation data. Here we restrict our attention to the case when the particles have spent a sufficiently long time in the bulk so that memory of the particles initial conditions is lost. We therefore initialize the particles in the middle of the channel and assume that the channel half-width $L$ is large compared to the persistence length $\tau_p v_0$ ($\tau_p  = D_r^{-1}$ for ABPs, $\tau_p = \gamma^{-1}$ for RTPs).

\subsection{The distribution of incoming angles}

\begin{figure}
    \centering
    \includegraphics[width = 8.5cm]{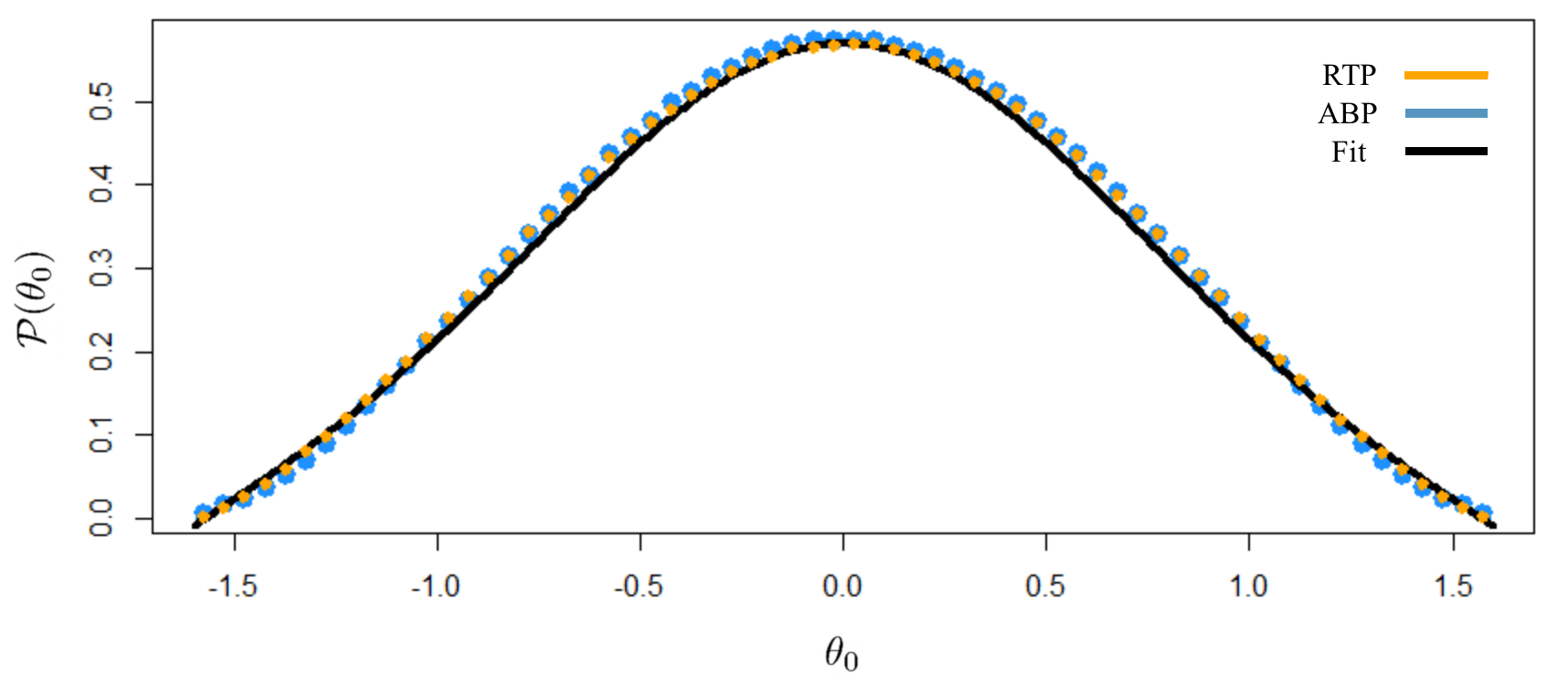}
    \caption{Plot showing the distribution of incoming angles $\mathcal{P}(\theta_0)$ for ABPs and RTPs. We see that the two particle species share the same distribution, with only very minor differences. Black line corresponds to best fit of the two first Fourier modes as described in the text.  }
    \label{fig:4}
\end{figure}

For ABPs originating far in the bulk hitting a flat boundary for the first time the distribution of incoming angles $\mathcal{P}(\theta_0 )$ is expected to be maximal at $\theta_0 = 0$ and decay monotonically towards zero at $\theta_0  = \pm \pi/2$, as eluded to in the previous section. The monotonic decay is expected since the larger the particles velocity component parallel to the boundary is, the more time noise has to act on its direction. Imagine a ABP close to a flat boundary. If noise reorients the particle away from the wall there is no incoming angle to speak of, while if the noise reorients the particles to move towards the wall, the incoming angle would be closer to $0$. Hence it is most likely to find incoming angles close to $0$. The same holds for RTPs, and the distribution of incoming angles is identical as seen in Fig. (\ref{fig:4}).

From our derivations we know that the RTP case is insensitive to the incoming angle, while in the ABP case the distribution of incoming angles will affect the FPT distribution. In this case it will be useful to observe that $\mathcal{P}(\theta_0)$ is well-approximated by the two first Fourier modes, $\mathcal{P}(\theta_0)$ is well approximated by the first couple of terms in its Fourier series
\begin{equation}
    \mathcal{P}(\theta_0) \approx \tilde{\mathcal{P}}_0 \cos(\theta_0) + \tilde{\mathcal{P}}_1 \cos(3 \theta_0)
\end{equation}
Normalization gives the constraint $\tilde{\mathcal{P}}_0 = (3 + 2 \tilde{\mathcal{P}}_1)/6$. From the numerical data, we find that $(\mathcal{P}_0,\mathcal{P}_1) = (0.508,0.061)$ best fits the data, as shown in Fig. (\ref{fig:4}). 

\subsection{Trapping statistics}

\begin{figure}[t]
    \centering
    \includegraphics[width = 8.5cm]{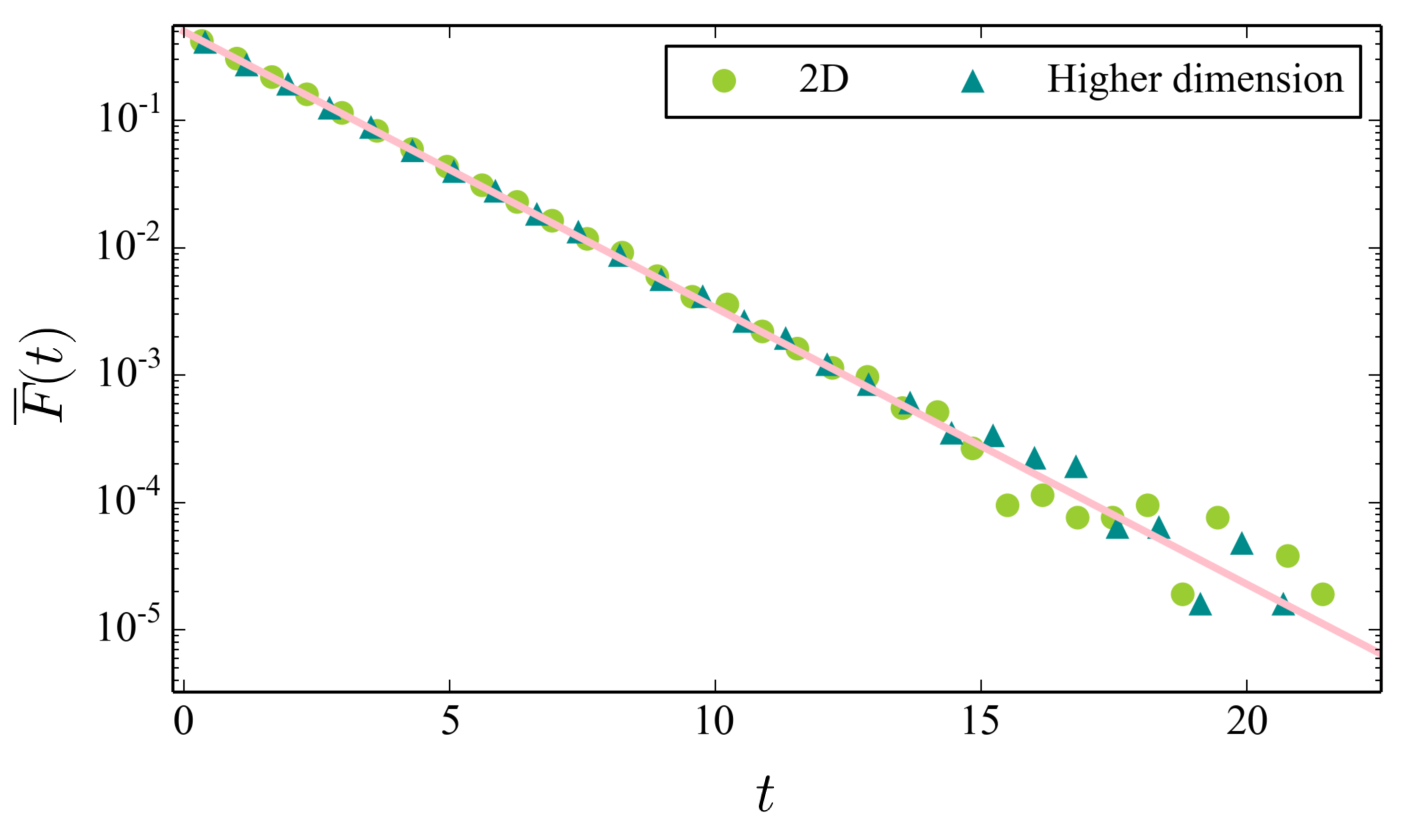}
    \caption{Plot showing first-passage time distribution for the run-and-tumble case, in units of the persistence time. Green circles result from numerical simulations of a 2D RTP as described in the dynamical equations. The dark blue triangles result from the simulation scheme of higher dimensional RTP as described in the text. As predicted the trapping statistics is universal for all dimensions higher than $1$, as shown by the pink solid line. }
    \label{fig:5}
\end{figure}

\textbf{RTP case:} As we have argued in previous sections, the run-and-tumble case has an universal exponentially decaying first passage time distribution  in any dimensions higher than one, taking the form
\begin{equation}
    \overline{F} (t) = F(t|\theta_0) =  \frac{\gamma}{2} e^{-\gamma t / 2}
\end{equation}
with no dependence on the incoming angle. The mean FPT is given simply by determined by the tumbling rate $T_1 = \overline{T}_1 = 2/\gamma$. Fig (\ref{fig:5}) shows the excellent agreement with this prediction of the numerical data for the trapping of run-and-tumble particles based on simulations of the dynamical equations Eq. (\ref{eq:dyneqs}). The higher-dimensional data was obtained as outlined in the end on section \ref{sec:rtp}.

\begin{figure}[t ]
    \centering
    \includegraphics[width = 8.5cm]{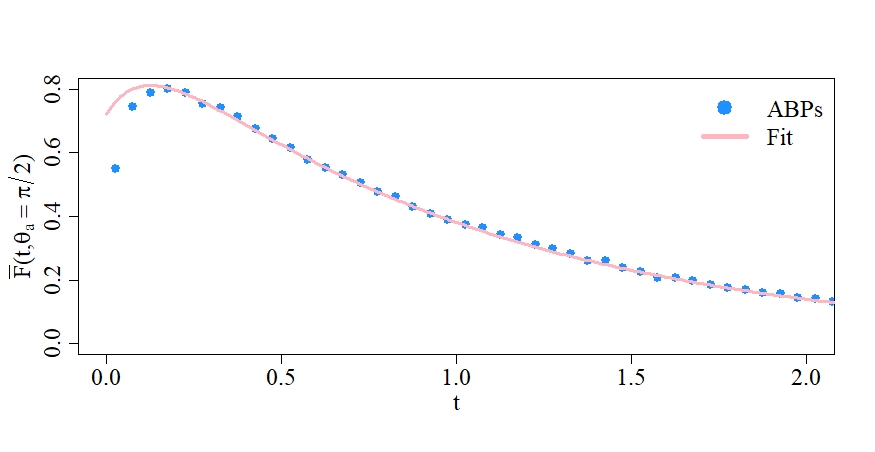}
    \caption{Trapping time distribution for ABPs in units of the persistence time, obtained from simulations (dots) and semi-analytical theory (solid line) based on numerically obtained parameters $\mathcal{P}_{0,1}$ for the distribution of incoming angles.  }
    \label{fig:6}
\end{figure}

\textbf{ABP case:}   Using Eq. (\ref{eq:avgFPTD}) for $\theta_a = \pi/2$ and keeping only the first two terms gives the collision-averaged FPT distribution
\begin{equation}
    \frac{1}{D_r}\overline{F}\left(t,\frac{\pi}{2}\right) =  \frac{3 + 2 \tilde{\mathcal{P}}_1}{3}e^{-D_r t} - 6  \tilde{\mathcal{P}}_1 e^{-9 D_r t}
\end{equation}
If the distribution of incoming angles was a perfect cosine ($\tilde{\mathcal{P}}_1 = 0$) the trapping time distribution would simply be a pure exponential and the mean trapping time the persistence time. Including the minor correction due to the second Fourier mode,we find the mean $\overline{T}_1 \approx 1.036$. While this corrections makes little difference to the numerical value of the mean trapping time, we see a clear effect of it in the trapping time distribution in Fig. (\ref{fig:6}), where the higher order terms are responsible for the non-monotonic behavior of the distribution for short trapping times. In other words, the likelihood of an ABP quickly escaping away from the wall is highly sensitive to its incoming angle. By contract, particles that resides at the wall longer escape with a probability that is independent of the incoming angle and falls off exponentially with residence time. 

\section{Discussion and outlook }\label{sec:outlook}

As summary, we have studied the trapping statistics of active Brownian and run-and-tumble particle at flat solid boundaries. Since the escape from the wall is noise-induced, we solve the first-passage problem associated with the angular dynamics of each particle type and find exact analytical expressions for the trapping time distribution. In the run-and-tumble case, the nature of the tumbling statistics makes the trapping time distribution independent of the incoming collision angle, while for the ABS case, the distribution depends strongly on the collision angle for short trapping times. We observe that both kinds of active particles share the same distribution of incoming angles, which is very well-approximated by its two first cosine modes. This was used to calculate semi-analytically the trapping time distribution averaged over incoming angles. The theoretical prediction agrees well with numerical simulations. 

Several striking differences between the two species is observed. First, the trapping time distribution of ABPs is non-monotonic for short trapping times, while the RTP distribution is purely exponential. Furthermore, we argue that the obtained distribution in the RTP case is universal for any spatial dimension, which we verified for the three dimensional case. The mean trapping duration for the ABP case is found to be close to 1.036 in units of the persistence time, while in the RTP case it is exactly $2$. While it is well-known that the dynamics of these two particle species are macroscopically equivalent in open spaces when using the same persistence time, we here show a simple example of a discrepancy between the two models in the presence of boundaries. 

While the distribution of incoming angles was found to be identical for both particle species, one should note that this will not be the case for the outgoing angle. In the RTP case, the escape angle is random and uniformly distributed, while ABPs will always leave the wall moving parallel to it. Throughout our analysis we have assumed that the particles are initialized sufficiently far from the wall so that any memory of initial conditions are lost. However, memory effects associated with multiple closely following trapping events is a case for further study, which may shed more light of similarities or differences between ABPs and RTPs trapped at walls. In this study, we have neglected alignment interactions between active particles as well as any hydrodynamic effects, which are important for the wall accumulation phenomenon. Thus, it would be interesting to study further their role on the wall-trapping statistics.

\begin{acknowledgements}
E.Q.Z.M., J.R and L.A. acknowledge support from the Research Council of Norway through the Center of Excellence funding scheme, Project No. 262644 (PoreLab). K.S.O acknowledges support from the Nordita Fellowship program. Nordita is partially supported by Nordforsk.
\end{acknowledgements}

\section*{References}

 \bibliographystyle{apsrev4-2}

%

\end{document}